\documentclass[conference]{IEEEtran}
\ifCLASSINFOpdf
\else
\fi
\usepackage{amsmath}
\usepackage{bm}
\usepackage{amssymb}
\usepackage{subcaption}
\usepackage{hyperref}
\usepackage[russian,english]{babel}
\usepackage{graphicx}
\graphicspath{ {./figures/} }
\newcommand{\Fig}[1]{Fig.~\textup{\ref{#1}}}
\usepackage[usenames,dvipsnames]{color}

\newcommand{\RomanNumeralCaps}[1]
    {\MakeUppercase{\romannumeral #1}}

\IEEEoverridecommandlockouts

\begin{document}
%



\title{Physical Modelling and Cancellation of External Passive Intermodulation in FDD MIMO
\thanks{The research was carried out at Skoltech and supported by the Russian Science Foundation (project no. 24-29-00189).}}


\author{
\centering\IEEEauthorblockN{1\textsuperscript{st} Stanislav Krikunov}
\IEEEauthorblockA{Skoltech \\
Moscow, Russia\\
Stanislav.Krikunov@skoltech.ru}
\and
\IEEEauthorblockN{2\textsuperscript{nd} Viacheslav Zemlyakov}
\IEEEauthorblockA{Southern Federal University\\
Rostov-on-Don, Russia\\
vvzemlyakov@sfedu.ru}
\and
\IEEEauthorblockN{3\textsuperscript{rd} Andrey Ivanov}
\IEEEauthorblockA{Skoltech \\
Moscow, Russia\\
AN.Ivanov@skoltech.ru}
\and}


%


\maketitle

\begin{abstract}
In this paper, the physical approach to model external (air-induced) passive intermodulation (PIM) is presented in a frequency-division duplexing (FDD) multiple-input multiple-output (MIMO) system with an arbitrary number of transceiver chains. The external PIM is a special case of intermodulation distortion (IMD), mainly generated by metallic objects possessing nonlinear properties (“rusty bolt” effect). Typically, such sources are located in the near-field or transition region of the antenna array. PIM products may fall into the receiver band of the FDD system, negatively affecting the uplink signal. In contrast to other works, this one directly simulates the physical external PIM. The system includes models of a point-source external PIM, a finite-length dipole antenna, a MIMO antenna array, and a baseband multicarrier 5G NR OFDM signal. The Channel coefficients method for multi-PIM-source compensation is replicated to verify the proposed external PIM modelling approach. Simulation results of artificially generated PIM cancellation show similar performance as real-life experiments. Therefore, the proposed approach allows testing PIM compensation algorithms on large systems with many antennas and arbitrary array structures. This eliminates the need for experiments with real hardware at the development stage of the PIM cancellation algorithm.
\end{abstract}


%
\IEEEpeerreviewmaketitle

\section{Introduction}

Enhanced Mobile Broadband (EMBB) is a service defined by the 3rd Generation Partnership Project for 4G Long-Term Evolution (LTE) and 5G New Radio (NR) deployment to provide higher data rates for the end user \cite{3GPP_38_913}. To achieve this, EMBB utilizes such technologies as Multiple Input Multiple Output (MIMO) \cite{3GPP_38_101, 3GPP_38_300}, Orthogonal Frequency Division Multiplexing (OFDM), and Carrier Aggregation (CA) \cite{PIM_OVERVIEW}. MIMO provides spatial signal diversity, OFDM provides frequency domain expansion, and CA allows flexible spectral resource allocation between different component carriers (CCs) of transmitted data \cite{5G_FUTURE}. In addition, LTE and NR specifications support the frequency division duplex (FDD) regime, where the transmitter (TX) and receiver (RX) operate simultaneously, occupying different frequency bands \cite{PIM_4}. However, real base station (BS) hardware is non-ideal and has nonlinear properties \cite{Clustering_Fitting, PAPR_STR, LUT_HPA, PA_NL}. This is especially noticeable when non-contiguously aggregated downlink (DL) signals pass through shared nonlinearities, the intermodulation products are generated \cite{PIM_2}. Some of them may fall into the RX band of the FDD system. All FDD transceivers have a duplexer between TX and RX chains, which protects the RX chain from intermodulation at the same frequency as TX CC. However, IMD products of CC interaction may fall into the RX band. Additionally, these products and products at other frequencies may affect surrounding systems working in frequency/time division duplexing modes as external sources of interference.

\begin{figure}[t]
    \centering
    \includegraphics[width=0.99\columnwidth]{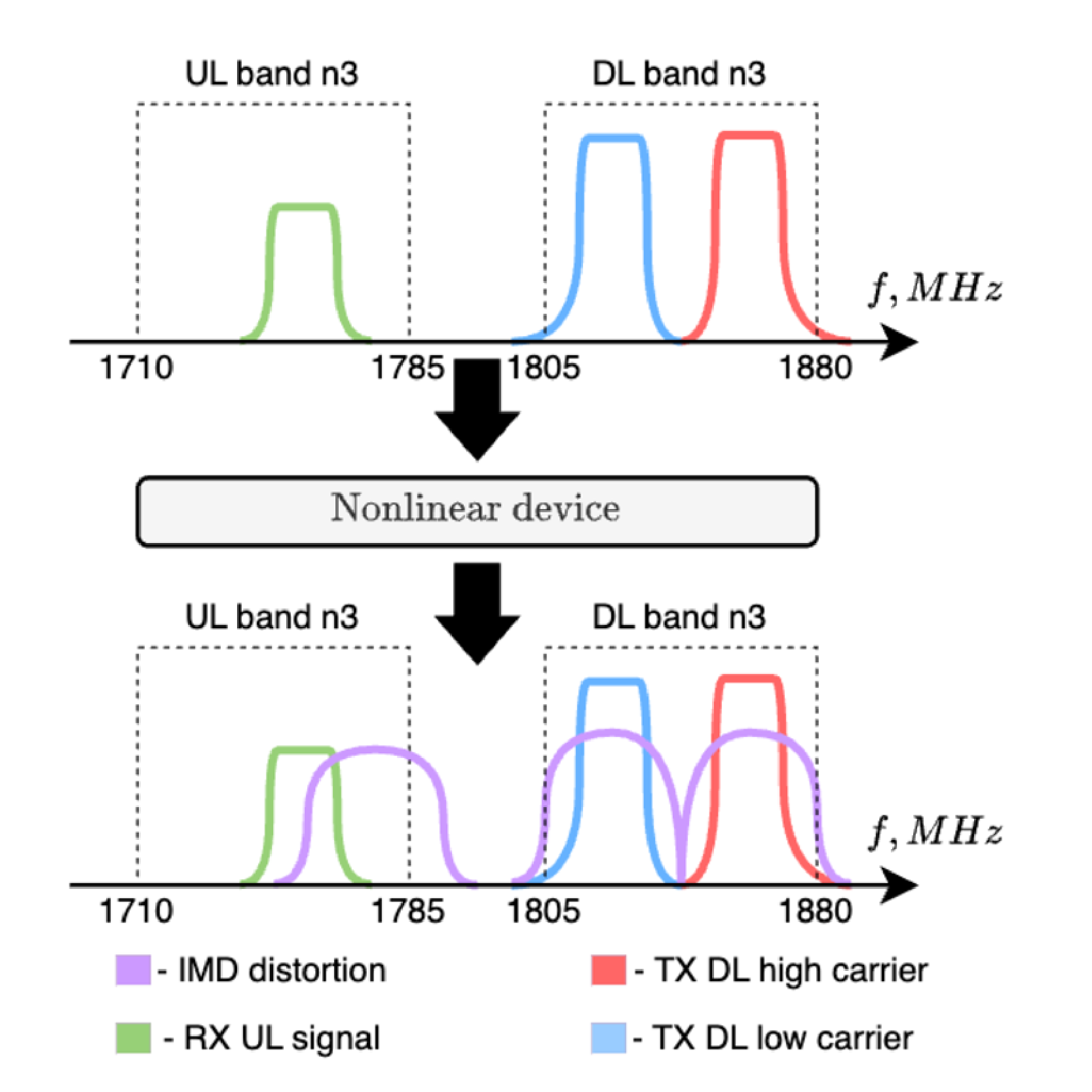}
    \caption{IMD-3 generation within n3 operating band.}
    \label{fig: PIM_generation}
\end{figure}

Meanwhile, nonlinear devices can be either passive or active. IMD products of passive devices are called Passive Intermodulation (PIM). Among PIM sources are weak mechanical connections in the TX chain, kinks and sharp edges in conductors, duplexer filters or ferrite fillers, switches, metal oxide layers covering conductors and junctions, and dirt in connectors. Passive nonlinearity is mainly caused by nonlinear conductive and magnetic properties of devices inside the transceiver chain (internal PIM) or outside the system in the near-field \cite{9976631} or transition antenna array region (external PIM, air-induced PIM) due to metal fences or billboards near the antenna array \cite{PIM_OVERVIEW}.

PIM represents one of the major interference problems \cite{10016907, 10167957, 9448862} in modern radio systems for service providers and equipment suppliers. PIM interference results in decreased coverage of BS cells, a decrease in the sensitivity of RX uplink (UL) signals, or possibly the complete inoperable transmission link.

Internal PIM can be eliminated by improving the production process of the equipment at the cost of manufacturing expenses. Unfortunately, external PIM cannot be eliminated in this way since it may be a part of a non-controllable built environment around the transceiver antenna system.


\subsection{Existing approaches}

Despite the wide variety of PIM compensation methods \cite{PIM_OVERVIEW, DIGITAL_CANCELLATION, PIM_NF_BF, ADAPTIVE_BEAMFORMING, IMD2CANCELLATION}, modern approaches do not allow one to simulate external PIM in the MIMO system physically. All known papers use a compensation model approach. In this case, the compensation model design follows the physical mechanism of external PIM generation with significant simplifications. Therefore, these research results are limited by real data measurements. Such kinds of measurement are not available for many researchers. Also, none of these methods provides a comprehensive process by which artificial interference caused by an external PIM source could be simulated directly, especially in an arbitrary MIMO system.
The authors of \cite{PIM_1, PIM_2, PIM_3, PIM_4, PIM_5} also mention that they are unaware of any works explicitly aiming to model and cancel air-induced PIM in FDD MIMO scenarios. 

It is also worth noting that in almost all recent articles, external PIM compensation methods intended for large-scale MIMO systems are tested only on base stations with 2 transceiver paths. Thus, works \cite{PIM_4, PIM_5} limit their research by a dual TX/RX chain MIMO system and 3 external sources of PIM. This is not enough for a comprehensive MIMO system test. Some works \cite{PIM_NF_BF} consider more complex systems with more transceiver chains. Unfortunately, the results of such works are difficult to reproduce since no data and simulation codes are available in open sources.

Consequently, developing a unified physical model that allows a realistic simulation of the external PIM phenomenon in FDD MIMO systems is an important direction from both theoretical and practical points of view.

\subsection{Contributions and Novelty}

All the external PIM cancellation algorithms mentioned in the literature require real hardware-measured testing data. Accessing the real BS hardware or measured data is impossible in most cases. Additionally, setting up an experiment and environment with an artificial PIM source is a rather expensive procedure requiring many specialists and additional preparations.

This paper presents a new physical model of artificial external PIM generation based on electromagnetic theory. The model has the following features:

1) The ability to generate any number of external point PIM sources based on arbitrary Uniform Rectangular Array (URA) structure;

2) The ability of near- or far-field zone effects and polarization effects to be considered due to the near-field dipole antenna model;

3) The ability to test both UL and DL PIM compensation methods;

4) A relatively simple model that does not require large computational resources and provides acceptable modelling accuracy;

\section{System model}

\begin{figure*}[ht]
    \centering
    \includegraphics[width=0.99\textwidth]{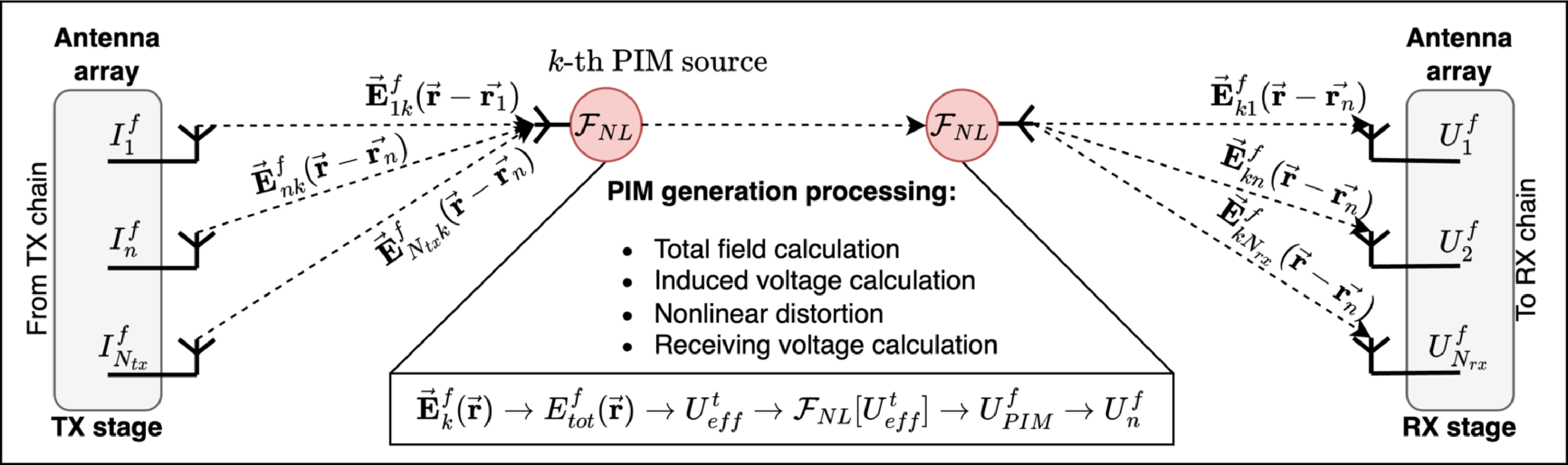}
    \caption{External PIM generation procedure.}
    \label{PIMdrawio}
\end{figure*}

In this paper, we apply Standard equivalent complex baseband signal modelling \cite{BB_MODELLING}. The process of generating PIM consists of the following steps:
\begin{enumerate}
\item OFDM signal precoding via the Discrete Fourier Transform Type \RomanNumeralCaps{1} codebook (designed for single-user MIMO mainly).
\item The frequency domain estimation of the electric field magnitude induced at an observation point by all antenna elements (external PIM source location).
\item A nonlinear element excitation and generation of passive intermodulation harmonics.
\item Backward propagation of intermodulation products at different frequencies compared to forward signal propagation.
\item Thermal noise distortion at the receiver chain. Duplexer filters are assumed to be ideal band-pass filters.
\end{enumerate}
The PIM generation process is illustrated in \Fig{PIMdrawio}.

\subsection{Single antenna model}

This work is based on a finite-length dipole antenna model of a zero-radius wire \cite{balanis2015antenna}. This allows for avoiding significant computational resource usage and provides good numerical agreement with a real antenna. Effects associated with the influence of magnetic fields are assumed to be negligible and are not considered in this article. For a dipole antenna oriented along the $z$-axis, the electric field at each frequency is given in cylindrical coordinates \eqref{eq: finite_length_dipole_field}. A sinusoidal current distribution is assumed.
\begin{equation}
 \begin{cases}
    E_{\rho}^f = j  \frac{\eta I l}{4 \pi \rho} \left[\Delta z_{-} G_1 + \Delta z_{+} G_2 - 2z \cos{(\frac{k_f l}{2})} G_0 \right] \\
    E_{\phi}^f = 0 \\
    E_{z}^f = -j \frac{\eta I l}{4 \pi} \left[G_1 + G_2 - 2 \cos{(\frac{k_f l}{2})} G_0 \right] \\
 \end{cases}
\label{eq: finite_length_dipole_field}
\end{equation}

\begin{flalign}
& G_0 = \frac{e^{-j k_f R_0}}{R_0}, G_1 = \frac{e^{-j k_f R_1}}{R_1}, G_2 = \frac{e^{-j k_f R_2}}{R_2} \nonumber & \\ 
& R_0 = \sqrt{\rho^2 + z^2}, R_1 = \sqrt{\rho^2 + \Delta z_{-}^2}, R_2 = \sqrt{\rho^2 + \Delta z_{+}^2} \nonumber & \\ 
& \rho = x^2 + y^2, \Delta z_{-} = z - \frac{l}{2}, \Delta z_{+} = z + \frac{l}{2} \nonumber,
\end{flalign}
\text{where} $x, y, z$ are the observation point coordinates in the cartesian coordinate system, $E_{\rho}, E_{\phi}, E_{z}$ are electric field components in the cylindrical coordinate system, $r$ is observation point distance, $\eta$ is free space impedance, $I$ is the complex amplitude of the current feeding the antenna and $l$ is antenna length. The average current feeding $n$-th antenna can be calculated using the Ohm law:
\begin{equation}
\overline{I}_n = \sqrt{\frac{P_n}{Z_n}},
\label{eq: ohm}
\end{equation}
\text{where} $P_n$ is antenna element radiating power, and $Z_n$ is antenna impedance. All antenna impedances are assumed to be the same, and antennas are considered not to affect each other. All antennas have the same radiating power.

Thus, the instantaneous value, if the input signal can be calculated as follows:

\begin{equation}
I_n = I_n(t) = \overline{I}_n \bm{U}_{n}(t)
\end{equation}

It is worth noting that the normalization to maintain power ratios is done as $\mathbb{E}[|\bm{U}_{n}(t)|^2]=1$ (for each TX chain separately).

\subsection{Antenna field coordinate transformation}
The following approach with coordinate transformation has been applied to model a dipole oriented along other axes (crossed dipoles in the x-y plane). $x,y,z$ are observation point coordinates associated with $\vec{\mathbf{r}}$ vector, $x_0^i,y_0^i,z_0^i$ are $n$-th antenna coordinates from the array associated with the $\vec{\mathbf{r}}_n$ vector. The initial dipole antenna is aligned along the $z$-axis. The electric field calculated in cylindrical coordinate system at the observation point $\vec{\mathbf{r}}_n(x-x_0^i,y-y_0^i,z-z_0^i)$ is $\vec{\mathbf{E}}(\vec{\mathbf{r}}-\vec{\mathbf{r}}_n)^{cyl}=\vec{\mathbf{E}}(\Delta \vec{\mathbf{r}}_n)^{cyl}$. Suppose that the dipole has been rotated in a local coordinate system under a linear transformation effect: $\mathbf{\Phi}_{coord}=\mathbf{\Phi}_{coord}(\Delta\vec{\mathbf{r}}_n)$ without changes in the observation point. In this case, the observation point field will be calculated as $\vec{\mathbf{E}}(\Delta\vec{\mathbf{r}}_n^*)^{cyl}=\vec{\mathbf{E}}(\mathbf{\Phi}_{coord}(\Delta\vec{\mathbf{r}}_n))^{cyl}$. At the same time, this field can be transformed into Cartesian coordinate system using linear transformation for vectors $\mathbf{\Phi}_{vect}=\mathbf{\Phi}_{vect}(\vec{\mathbf{r}}_n^*)$: $\vec{\mathbf{E}}(\vec{\mathbf{r}}_n^*)^{cart} = \mathbf{\Phi}_{vect} \vec{\mathbf{E}}(\Delta\vec{\mathbf{r}}_n^*)^{cyl}$.
Inverse coordinate transformation can be used to obtain the field in the original coordinate system: $\vec{\mathbf{E}}(\vec{\mathbf{r}}_n)^{cart} = \mathbf{\Phi}_{coord}^{-1}\vec{\mathbf{E}}(\Delta\vec{\mathbf{r}}_n^*)^{cart}$.

\subsection{Antenna array model}

A typical MIMO antenna array consists of rows and columns of dual-polarized antenna elements radiating from the same point (crossed dipoles to consider polarisation properties). The whole array can be divided into subarrays. Each subarray is connected to two radio chains, normally one per polarization. Antennas within the subarray utilize tunable phase shifts to provide electric antenna directivity pattern tilt.

In practice, even if the antenna array consists of identical antenna elements, their current distributions may differ from the law of current distribution in a separate antenna due to the mutual coupling effect. In the current work, the mutual coupling effect is assumed to be negligible. Assume the current density distribution on the $n$-th antenna element as

\begin{equation}
\vec{\bm J_n} = I_n \vec{\bm J} (\vec{\bm{r}}-\vec{\bm{r}}_n),
\end{equation}
\text{where} $I_n$ is the complex amplitude of the current applied to the $n$-th antenna element,  $\vec{\bm{J}}$ is current density distribution in a local coordinate system of the antenna, $\vec{\bm {r}}$ and $\vec{\bm{r_n}}$ are observation point radius vectors and $n$-th antenna position in a global coordinate system, respectively. This current density distribution produces an electric field $\vec{\bm E}_n^{f}[\vec{\bm J_n}]$ at the observation point at frequency $f$, which can be calculated using \eqref{eq: finite_length_dipole_field}.

Thus, the total electric field $\vec{\bm {E}^{f}}$ produced by $N_{TX}$ antenna elements at the observation point $\vec{\bm {r}}$ can be represented as a superposition of the fields of individual antenna elements, taking into account the exciting currents:

\begin{equation}
\vec{\bm E}^{f}(\vec{\bm{r}}) = \sum_{n=0}^{N_{TX}-1} I_n^{f} \vec{\bm E}_n^{f}[\vec{\bm J_n}(\vec{\bm{r}}-\vec{\bm{r}}_n)]
\label{eq: total_array_field}
\end{equation}

\subsection{Point source model of external PIM}
Assume that the electric field at the PIM source location (observation point) is $\vec{\bm E}^{f}(\vec{\bm{r}})$ is calculated using \eqref{eq: total_array_field}. The total field exciting the PIM source can be represented as a product of PIM source orientation vector $\vec{\bm{p}}=\vec{\bm{p}}(p_x,p_y,p_z), |\vec{\bm{p}}|=1$ and electric field vector at the given point:

\begin{equation}
E_{tot}^{f}(\vec{\bm{r}}) = \vec{\bm E}^{f}(\vec{\bm{r}}) \cdot  \vec{\bm{p}}
\end{equation}

The induced voltage at the observation point is proportional to the electric field:

\begin{equation}
U_{eff}^{f} \propto E_{tot}^{f}(\vec{\bm{r}})
\end{equation}

The voltage after nonlinear distortion is proportional to a non-linear function of $U_{eff}^{t}=\operatorname{IDFT}(\sum_f U_{eff}^{f})$ taken in the time domain:

\begin{equation}
U_{PIM}^f = \operatorname{DFT}(\mathcal{F}_{NL} [U_{eff}^{t}]),
\end{equation}
\text{where} $\operatorname{DFT}$ and $\operatorname{IDFT}$ are direct and inverse Discrete Fourier Transforms respectively. The radiation field in the reverse direction can be represented similarly to the forward path. In this case, the signal at $n$-th receiving antenna terminal will be defined as:

\begin{equation}
U_{n}^{f} \propto U_{PIM}^f \vec{\bm E}_{n}^{f}(\vec{\bm{r}})
\end{equation}
Assume the level of PIM generated by a point source is specified by normalization relative to the level of thermal noise. This does not affect the amplitude and phase relationship between the generating signals and the generated PIM. This is because, physically, a point source emits a tiny amount of energy, which cannot be considered using the point source model.

\subsection{Nonlinear element model}
Nonlinear elements, depending on their nature, may also have memory effects \cite{LUT_HPA}. In this case, a nonlinear element producing IMD harmonics can be modelled in various ways. One common approach is Generalized Memory Polynomial (GMP) \cite{GMP}. Such nonlinear function $\mathcal{F}_{NL}=\mathcal{F}_{NL} [\mathcal{M},\mathcal{K},\mathcal{P}]$ has the following form of \eqref{eq: nonlinearity}.
\begin{equation}
\mathcal{F}_{NL} = \sum_{m\in{\mathcal{M}}}\sum_{k \in \mathcal{K}} \sum_{p=1}^{\mathcal{P}} g_{mkp}u(t-m)|u(t-m-k)|^{p-1},
\label{eq: nonlinearity}
\end{equation}
\text{where} $\mathcal{M}$ is the number of global precursor ($\mathcal{M}<0$) and postcursor ($\mathcal{M}>0$) taps, respectively, $\mathcal{K}$ - the number of envelope lead ($\mathcal{K}<0$) and lag ($\mathcal{K}>0$) samples; $\mathcal{P}$ is polynomial order, $g$ is amplitude coefficient, $u$ is input signal, $t$ is time index.

\begin{figure}[t]
    \centering
    \includegraphics[width=0.99\columnwidth]{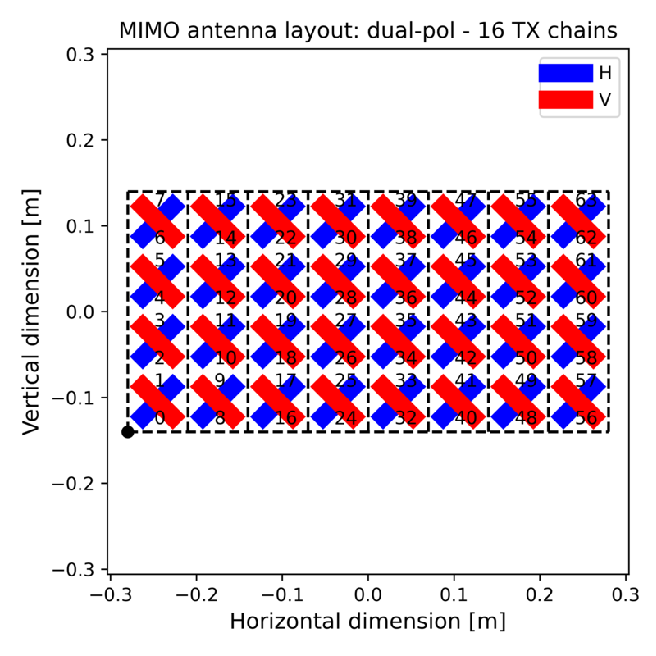}
    \caption{16 TX/RX-chain antenna structure.}
    \label{fig: ANTENNA_1PP_16T16R_5MHZ}
\end{figure}

\begin{table}[]
\centering
\begin{tabular}{|l|c|c|} 
\hline
\textbf{Parameter} & \textbf{Carrier bandwidth} \\ 
\hline
Bandwidth [MHz] & 5 \\ 
\hline
Low CC frequency [MHz] & 1819 \\ 
\hline
High CC frequency [MHz] & 1866.5 \\ 
\hline
BB sampling rate [MHz] & 7.68 \\ 
\hline
RF sampling rate [MHz] & 122.8 \\ 
\hline
Oversampling factor & 16 \\ 
\hline
FFT length [samples] & 512 \\ 
\hline
Training sequence length [samples] & 131072 \\ 
\hline
Testing sequence length [samples] & 65536 \\ 
\hline
Signal duration [ms] & 24.5 \\ 
\hline
\end{tabular}
\caption{OFDM signal simulation parameters}
\label{Tab:signal_parameters}
\end{table}

\section{Simulation results}
The compensation approach for model testing is based on the Channel coefficients method \cite{PIM_4}. Channel coefficients here are adaptive weights of different nonlinear terms taken into account. Upsampled low and high-frequency component carriers, referred to as basis functions (BFs) of each antenna signal, are weighted using carrier-wise adaptive channel coefficients. The original Levenberg–Marquardt optimization algorithm is replaced with Stochastic gradient descent (SGD) for faster implementation and better adaptation. The least squares (LS) coefficient estimation of BFs left unchanged. RX signal represents pure external PIM mixed with thermal noise. The algorithm tries to suppress external PIM at the level of thermal noise floor (NF).

\begin{figure}[t]
    \centering
    \includegraphics[width=0.99\columnwidth]{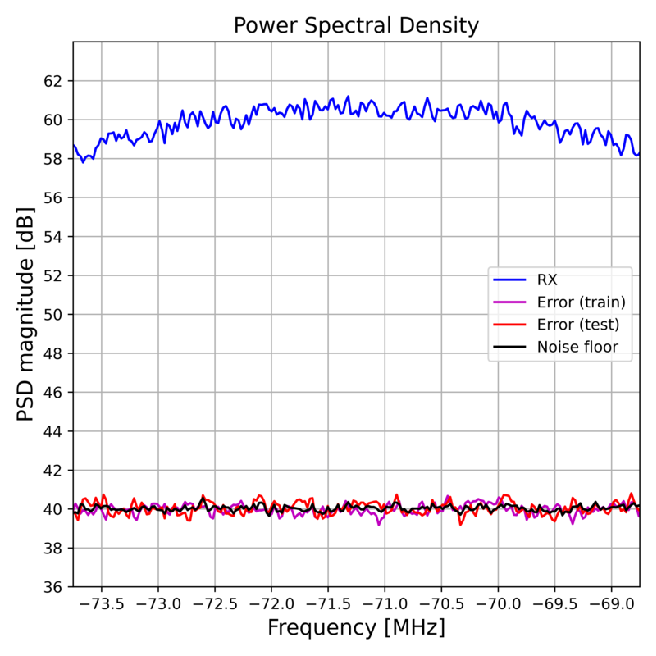}
    \caption{Cancellation of the 5-MHz carrier PIM from a single-point PIM source in 16T16R MIMO.}
    \label{fig: PSD_1PP_16T16R_5MHZ}
\end{figure}

\begin{figure}[t]
    \centering
    \includegraphics[width=0.99\columnwidth]{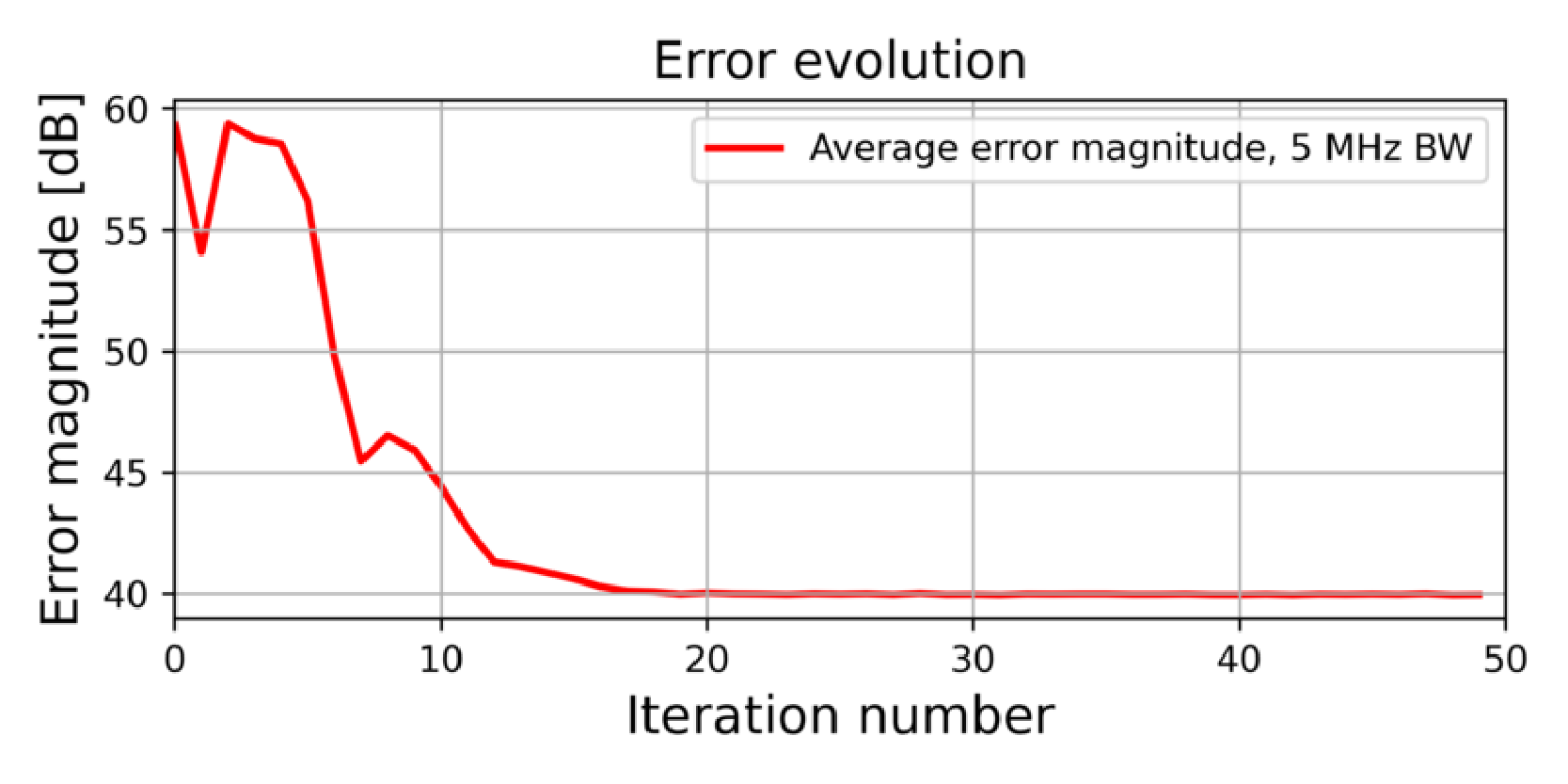}
    \caption{Convergence of the average cancelled RX PIM signal power in 16T16R MIMO from a single-point PIM source.}
    \label{fig: ERR_1PP_16T16R_5MHZ}
\end{figure}

\begin{figure}[t]
    \centering
    \includegraphics[width=0.99\columnwidth]{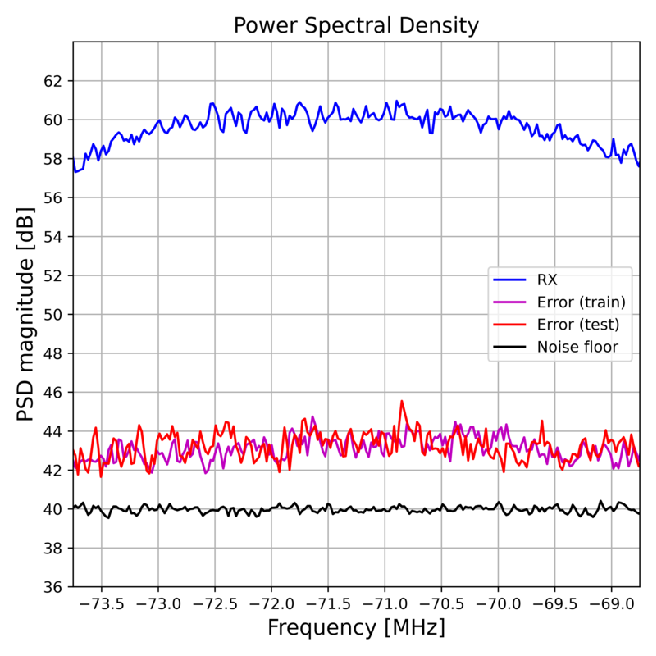}
    \caption{Cancellation of the 5-MHz carrier PIM from a three-point PIM source in 16T16R MIMO.}
    \label{fig: PSD_3PP_16T16R_5MHZ}
\end{figure}

\subsection{Simulation setup}
The modelling was carried out for 5G NR band n3 shown in \Fig{fig: PIM_generation}. The component CCs of 5MHz bandwidths are located at the 1819 MHz and 1866.5 MHz frequencies respectively and their intermodulation products lie at the 1771.5 MHz central frequency (-71.25 MHz w.r.t. TX zero frequency). In baseband digital processing, where the PIM cancellation occurs, the TX and RX signals are oversampled by a factor of 16 for the BF generation. Artificially generated data is divided into 2 sets: a training set of 131072 samples and a testing set of 65536 samples. Signal simulation parameters are summarized in Table \ref{Tab:signal_parameters}.

In this work, the nonlinearity for PIM generation follows the structure of \eqref{eq: nonlinearity} with the following parameters for simplicity: $\mathcal{P}=\{3\}, \mathcal{M}=\{0\}, \mathcal{K}=\{0\}$. Thus, a simple, memoryless, nonlinear model of IMD-3 is assumed. The nonlinear model of the compensation algorithm follows the same structure as well. However, more complex parameters are chosen under the assumption of unknown nonlinearity behaviour at the external PIM source: $\mathcal{P}=\{3,5\}, \mathcal{M}=\{-2,-1,0,1,2\}, \mathcal{K}=\{-1,0,1\}$. The orientation of the PIM source $\vec{\bm{p}}$ is random for each scenario. The total PIM level is artificially scaled to a reasonable value.

Each antenna in the simulations is assumed to be ideally matched to the transceiver path, has a 50 Ohm impedance, and has an average radiating power of 37 dBm. The conversion of energy into radiation occurs completely.

Two scenarios of external PIM source location are considered:

1) \textbf{Scenario 1:} A single external PIM source is located in front of the centre of the antenna array at a distance of 2.5 m relative to it. PIM cartesian coordinates $\vec{\bm{r}}$ are: \{0, 0, 2.5\}.

2) \textbf{Scenario 2:} Three external PIM sources are located in front of the antenna with a 1 m separation along the $x$-axis and a 0.5 m separation along the $y$-axis. PIM cartesian coordinates $\vec{\bm{r}}$ are: \{-1, -0.5, 3\}, \{0, 0, 3\}, \{1, 0.5, 3\}.

\subsection{External PIM compensation in 16T16R MIMO}
The result for a large-scale MIMO system with 16 transmitting and receiving antennas is shown. Each transceiver chain in the system is represented by a set of half-wavelength dipoles grouped into unit cells of 4 antennas. The signal from each channel is divided equally among all antennas within a unit cell. The distance between the antennas is equal to half of the wavelength (w.r.t. the central frequency of the TX signal band). Even channels have vertical polarization; odd channels have horizontal polarization (crossed dipoles). The antenna layout is shown in \Fig{fig: ANTENNA_1PP_16T16R_5MHZ} where numbering means antenna number which is not related to the number of TX chains directly. A dashed line limits unit cells.

\subsubsection{Single point source compensation}
The compensation ability of the algorithm for \textbf{scenario 1} is shown in \Fig{fig: PSD_1PP_16T16R_5MHZ}. External PIM is cancelled to the noise floor after full algorithm convergence (\Fig{fig: ERR_1PP_16T16R_5MHZ}).
\subsubsection{Multiple point source compensation}
The compensation ability of the algorithm for \textbf{scenario 2} is shown in \Fig{fig: PSD_3PP_16T16R_5MHZ}. In this case, external PIM cannot be cancelled to the noise floor which can be caused when the optimization algorithm hits a local minimum.

\subsection{Near-field PIM power variation}

\begin{figure}
    \begin{minipage}{.48\textwidth}
    \centering
    \subfloat[]{\includegraphics[width=.45\linewidth]{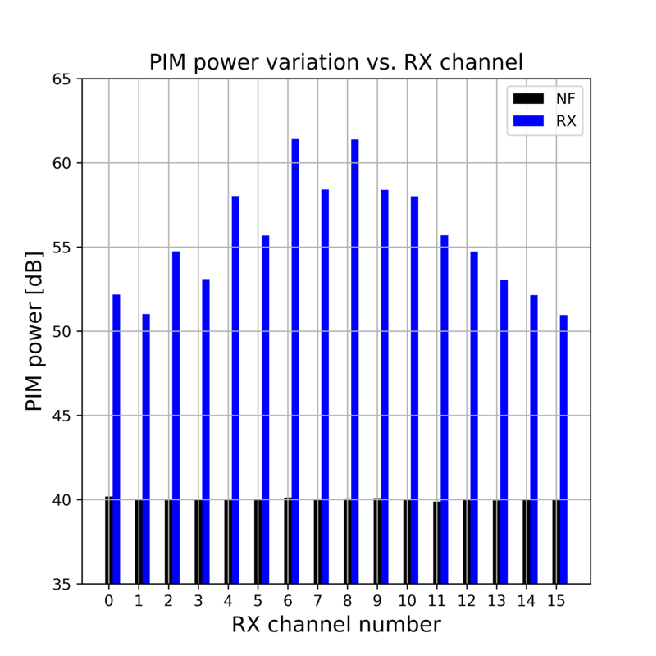}}\quad
    \subfloat[]{\includegraphics[width=.45\linewidth]{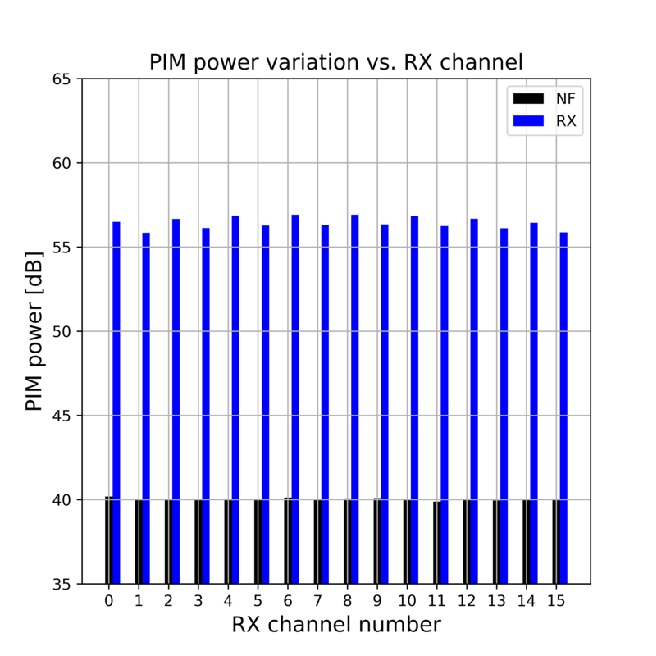}}
    \caption{RX PIM power variation at the distance of 0.1 m from the antenna array (left, high power variation) and 1m (right, low power variation).}
    \label{fig:testa}
    \end{minipage}\hfill
    \end{figure}

The location of the PIM source significantly affects the received power distribution between the channels. This is especially noticeable in the near-field zone of the antenna array since the wave here is spherical and the electric field varies a lot. As the source moves away from the antenna array, the wave becomes flat, and such a power variation is no longer noticeable (\Fig{fig:testa}). Polarization effects can also be noticed since the signal power at even antennas differs from the signal of odd antennas.

\section{Conclusion}
This paper presents an approach to modelling external sources of passive intermodulation in 5G FDD MIMO systems. The approach allows for simulating PIM sources including the near-field zone of the MIMO antenna array and does not require computationally complex full-wave electromagnetic modelling. The external PIM effect can be reproduced with sufficient accuracy for PIM compensation algorithm testing on large-scale MIMO systems. The cancellation results of artificially generated PIM are aligned with experiments conducted on real data available from the literature. Thus, this approach can be used to test and debug algorithms to cancel PIM from external sources.

\small
\bibliographystyle{ieeetr}
\bibliography{REFERENCES}

\begin{thebibliography}{10}

\bibitem{3GPP_38_913}
3GPP, ``3rd generation partnership project; technical specification group radio access network; study on scenarios and requirements for next generation access technologies (release 17),'' Tech. Rep. TR 38.913-03, v17.0.0 (2022-03), 3GPP, 2022.

\bibitem{3GPP_38_101}
3GPP, ``3rd generation partnership project; technical specification group radio access network; user equipment (ue) radio transmission and reception; part 3: Range 1 and range 2 interworking operation with other radios (release 17),'' Tech. Rep. TR 38.101-03, v17.7.0 (2022-09), 3GPP, 2022.

\bibitem{3GPP_38_300}
3GPP, ``3rd generation partnership project; technical specification group radio access network; nr; nr and ng-ran overall description; stage 2 (release 17),'' Tech. Rep. TR 38.300, v16.1.0 (2022-09), 3GPP, 2022.

\bibitem{PIM_OVERVIEW}
T.~Ahmmed, A.~Kiayani, R.~M. Shubair, and H.~Yanikomeroglu, ``Overview of passive intermodulation in modern wireless networks: Concepts and cancellation techniques,'' {\em IEEE Access}, vol.~11, pp.~128337--128353, 2023.

\bibitem{5G_FUTURE}
W.~Chen, X.~Lin, J.~Lee, A.~Toskala, S.~Sun, C.~F. Chiasserini, and L.~Liu, ``5g-advanced toward 6g: Past, present, and future,'' {\em IEEE Journal on Selected Areas in Communications}, vol.~41, no.~6, pp.~1592--1619, 2023.

\bibitem{PIM_4}
V.~Lampu, L.~Anttila, M.~Turunen, M.~Fleischer, J.~Hellmann, and M.~Valkama, ``Air-induced passive intermodulation in fdd mimo systems: Algorithms and measurements,'' {\em IEEE Transactions on Microwave Theory and Techniques}, vol.~71, no.~1, pp.~373--388, 2023.

\bibitem{Clustering_Fitting}
S.~Krikunov, R.~Bychkov, A.~Blagodarnyi, and A.~Ivanov, ``Clustering and fitting to reduce papr in multi-user ofdm systems,'' in {\em 2023 25th International Conference on Digital Signal Processing and its Applications (DSPA)}, pp.~1--6, 2023.

\bibitem{PAPR_STR}
A.~Ivanov and D.~Lakontsev, ``Selective tone reservation for papr reduction in wireless communication systems,'' in {\em 2017 IEEE International Workshop on Signal Processing Systems (SiPS)}, pp.~1--6, 2017.

\bibitem{LUT_HPA}
A.~Ivanov and D.~Lakontsev, ``Adaptable look-up tables for linearizing high power amplifiers,'' in {\em 2017 3rd International Conference on Frontiers of Signal Processing (ICFSP)}, pp.~96--100, 2017.

\bibitem{PA_NL}
P.~Plotnikov and D.~Dolgikh, ``Joint compensation of power amplifier nonlinear distortions on transmitter and receiver sides,'' in {\em 2019 IEEE International Black Sea Conference on Communications and Networking (BlackSeaCom)}, pp.~1--5, 2019.

\bibitem{PIM_2}
V.~Lampu, L.~Anttila, M.~Turunen, M.~Fleischer, J.~Hellmann, and M.~Valkama, ``Air-induced pim cancellation in fdd mimo transceivers,'' {\em IEEE Microwave and Wireless Components Letters}, vol.~32, no.~6, pp.~780--783, 2022.

\bibitem{9976631}
V.~Abramian and A.~Larionov, ``Numerical research of the probability of radio frequency identification of tags using a uav-mounted rfid reader,'' in {\em 2022 International Conference on Information, Control, and Communication Technologies (ICCT)}, pp.~1--5, 2022.

\bibitem{10016907}
A.~Blagodarnyi, R.~Bychkov, S.~Krikunov, and A.~Ivanov, ``Tensor-assisted cnn to estimate channel in massive mimo,'' in {\em 2022 IEEE International Multi-Conference on Engineering, Computer and Information Sciences (SIBIRCON)}, pp.~20--24, 2022.

\bibitem{10167957}
D.~Artemasov, A.~Blagodarnyi, A.~Sherstobitov, and V.~Lyashev, ``Vector autoregression model utilization for massive-mimo channel denoising,'' in {\em 2023 International Balkan Conference on Communications and Networking (BalkanCom)}, pp.~1--6, 2023.

\bibitem{9448862}
D.~Yarotsky, A.~Ivanov, R.~Bychkov, A.~Osinsky, A.~Savinov, M.~Trefilov, and V.~Lyashev, ``Machine learning-assisted channel estimation in massive mimo receiver,'' in {\em 2021 IEEE 93rd Vehicular Technology Conference (VTC2021-Spring)}, pp.~1--5, 2021.

\bibitem{DIGITAL_CANCELLATION}
M.~Z. Waheed, P.~P. Campo, D.~Korpi, A.~Kiayani, L.~Anttila, and M.~Valkama, ``Digital cancellation of passive intermodulation in fdd transceivers,'' in {\em 2018 52nd Asilomar Conference on Signals, Systems, and Computers}, pp.~1375--1381, 2018.

\bibitem{PIM_NF_BF}
Z.~Ye, X.~Zhu, X.~Zhang, and Y.~Zhang, ``External passive intermodulation suppression by near-field downlink beamforming,'' {\em IEEE Transactions on Microwave Theory and Techniques}, vol.~72, no.~4, pp.~2355--2367, 2024.

\bibitem{ADAPTIVE_BEAMFORMING}
Z.~Wang, X.~Zhu, Y.~Jiang, H.~Zeng, and B.~Li, ``External passive intermodulation suppression by general linear combination based robust adaptive beamforming,'' in {\em 2022 IEEE 96th Vehicular Technology Conference (VTC2022-Fall)}, pp.~1--6, 2022.

\bibitem{IMD2CANCELLATION}
A.~A. Degtyarev, N.~V. Bakholdin, A.~Y. Maslovskiy, and S.~A. Bakhurin, ``Brief research of traditional and ai-based models for imd2 cancellation,'' 2024.

\bibitem{PIM_1}
V.~Lampu, L.~Anttila, M.~Turunen, M.~Fleischer, J.~Hellmann, and M.~Valkama, ``Air-induced passive intermodulation in fdd networks: Modeling, cancellation and measurements,'' in {\em 2021 55th Asilomar Conference on Signals, Systems, and Computers}, pp.~983--988, 2021.

\bibitem{PIM_3}
M.~Z. Waheed, V.~Lampu, A.~Kiayani, M.~Fleischer, L.~Anttila, and M.~Valkama, ``Modeling and digital suppression of passive nonlinear distortion in simultaneous transmit—receive systems,'' in {\em 2022 56th Asilomar Conference on Signals, Systems, and Computers}, pp.~1339--1344, 2022.

\bibitem{PIM_5}
V.~Lampu, L.~Anttila, M.~Turunen, M.~Fleischer, J.~Hellmann, and M.~Valkama, ``Cancellation of air-induced passive intermodulation in fdd mimo systems: Low-complexity cascade model and measurements,'' in {\em 2023 IEEE/MTT-S International Microwave Symposium - IMS 2023}, pp.~33--36, 2023.

\bibitem{BB_MODELLING}
Y.~Ye, D.~Spina, D.~Deschrijver, W.~Bogaerts, and T.~Dhaene, ``Efficient time-domain modeling and simulation of passive bandpass systems,'' in {\em 2019 International Conference on Electromagnetics in Advanced Applications (ICEAA)}, pp.~0992--0996, 2019.

\bibitem{balanis2015antenna}
C.~Balanis, {\em Antenna Theory: Analysis and Design}.
\newblock Wiley, 2015.

\bibitem{GMP}
D.~Morgan, Z.~Ma, J.~Kim, M.~Zierdt, and J.~Pastalan, ``A generalized memory polynomial model for digital predistortion of rf power amplifiers,'' {\em IEEE Transactions on Signal Processing}, vol.~54, no.~10, pp.~3852--3860, 2006.

\end{thebibliography}

\end{document}